\documentclass[letterpaper, 10 pt, conference]{ieeeconf}  

\IEEEoverridecommandlockouts                              

\usepackage{mathptmx} 
\usepackage{times} 
\usepackage{amsmath}
\usepackage{amssymb} 
\usepackage{float}
\usepackage{subcaption} 
\usepackage{graphicx}
\usepackage[T1]{fontenc}

\usepackage{tikz}
\usetikzlibrary{shapes, arrows.meta, positioning, calc}
\usepackage{pgfplots}
\usepackage{graphicx} 

\usepackage{amsthm}
\theoremstyle{remark}
\newtheorem{remark}{Remark}
\usepackage{setspace}
\usepackage{xcolor}
\usepackage{amsmath}
\usepackage{amsfonts}
\usepackage{bbm}
\usepackage{mathtools} 
\RequirePackage{algorithm}
\RequirePackage{algpseudocode} 
\usepackage{caption}
\usepackage{subcaption}
\usepackage{comment}
\usepackage{listings}
\usepackage{dsfont}
\usepackage{amsfonts}
\usepackage{booktabs}
\usepackage{hyperref}
\usepackage{url}
\usepackage{bm}
\usepackage{dsfont}

\newcommand{\numherders}[0]{N}
\newcommand{\numtargets}[0]{M}

\newcommand{\goalradius}[0]{\rho_{\mathrm{G}}}

\newcommand{\samplingtime}[0]{\Delta t}

\usepackage{dsfont}

\newcommand{\B}[1]{\if#1\relax\bm{#1}\else\mathbf{#1}\fi} 

\title{\LARGE \bf
Multi-robot obstacle-aware shepherding of non-cohesive target agents }
\author{Cinzia Tomaselli\textsuperscript{1}, Stefano Covone\textsuperscript{1}, Andreagiovanni Reina\textsuperscript{2,3}, Mario di Bernardo\textsuperscript{1,4,*}
\thanks{\textsuperscript{1}Scuola Superiore Meridionale, Naples, Italy}%
\thanks{\textsuperscript{2}Centre for the Advanced Study of Collective Behaviour, Universit\"{a}t Konstanz, Konstanz, Germany}
\thanks{\textsuperscript{3}Department of Collective Behaviour, Max Planck Institute of Animal Behavior, Konstanz, Germany}%
\thanks{\textsuperscript{4}Department of Electrical Engineering and Information Technology, University of Naples Federico II, Naples, Italy}
\thanks{\textsuperscript{*} Corresponding author (mario.dibernardo@unina.it)}}

\begin{document}
\maketitle
\thispagestyle{empty}
\pagestyle{empty}

\begin{abstract}
This paper presents a novel control strategy for multi-agent shepherding of non-cohesive targets in obstacle-rich environments. Unlike previous approaches that assume cohesive flocking behavior, our method handles targets that interact only with nearby herders through repulsive forces and exhibit no inter-target coordination. Each herder employs a hybrid control policy that combines direct goal-oriented steering with obstacle-tangent maneuvering, enabling targets to circumnavigate obstacles while being guided toward a goal region. The herder dynamics integrate three key behaviors: return-to-goal motion when idle, target steering with adaptive directional control, and obstacle avoidance using both normal and tangential force components. Numerical simulations demonstrate superior performance compared to existing shepherding methods, achieving higher target confinement rates in cluttered environments. Experimental validation using TurtleBot4 herders and Osoyoo target robots in an indoor arena confirms the practical effectiveness of the proposed approach. 
\end{abstract}

\section{Introduction}
 Multi-agent shepherding represents a challenging distributed control problem where herder agents must coordinate to guide independently moving targets to desired spatial configurations \cite{long2020comprehensive}. This bio-inspired paradigm, observed in nature through examples like sheepdogs herding sheep \cite{strombomSolvingShepherdingProblem2014} or orcas coordinating to hunt fish \cite{haque2011biologically}, has attracted significant attention in robotics due to its broad application potential in crowd control, search-and-rescue operations, area defense, and autonomous vehicle coordination (see \cite{long2020comprehensive} for a comprehensive review of applications).
 
Unlike traditional formation control or consensus problems where all agents cooperate toward common objectives, shepherding involves non-cooperative targets that move independently according to their own dynamics \cite{muro_wolf-pack_2011}. This fundamental difference creates unique control challenges including heterogeneous agent populations, indirect influence mechanisms \cite{licitraSingleAgentIndirectherding2018}, and the need for real-time distributed decision-making without direct target communication \cite{li2023communication}.

The transition from abstract models to real-world applications introduces several additional challenges that must be addressed to ensure effective control. Among such challenges, the presence of obstacles in the environment plays a crucial role, as it can significantly affect the motion of agents, requiring control strategies to account for complex navigation scenarios.

Most existing shepherding strategies rely on heuristic coordination rules or model-based nonlinear or optimal control approaches, e.g., \cite{piersonControllingNoncooperativeHerds2018,licitraSingleAgentIndirectherding2018,zhang2024distributed}. A critical limitation is the common assumption that targets exhibit cohesive collective behavior (e.g., flocking), enabling the group to be treated as a single controllable entity \cite{lien2004shepherding}. However, cohesive behavior frequently fails in practical scenarios, requiring herders to influence each target individually and dramatically increasing control complexity \cite{aulettaHerdingStochasticAutonomous2022,lamaShepherdingControlHerdability2024,koAsymptoticBehaviorControl2020}.

This is also commonly exploited to drive targets around environmental obstacles by treating them as a cohesive group and aiming at controlling their centre of mass \cite{piersonControllingNoncooperativeHerds2018}.

In this work, we extend existing shepherding strategies to the case of non-cohesive targets moving in obstacle-rich environments. Specifically, we augment the basic repulsive herder-target interaction with a hybrid control strategy that -- inspired by \cite{deMedioRobot1991} -- combines repulsive and tangential forces around obstacle boundaries, enabling herders to guide targets around obstacles through circumnavigation maneuvers. This extension addresses the practical challenge of deploying shepherding algorithms in realistic environments where navigation constraints significantly impact system performance without requiring any assumption of cohesive behaviour in the targets.

\subsection{Related work}
In previous work, obstacles have been represented in shepherding problems either through explicit geometric descriptions, such as polygons with smoothed vertices or continuous boundary curves \cite{geopf2024}, or implicitly through repulsive potential fields that influence agents’ motion \cite{vanhavermaet2024reactive}. Given these representation choices, shepherding approaches must address obstacle avoidance through task-specific strategies that ensure both herder safety during navigation and effective target steering around environmental constraints.

Existing methods adopt different computational trade-offs to achieve these dual objectives, though most assume cohesive target behavior. Hybrid planning approaches prioritize herder safety through local reactive terms while using map-based planning to guide cohesive target groups toward the goal \cite{hamandi2024robotic}, achieving near-optimal global paths at significant computational expense. Similarly, other methods employ graph-based path planning for herder navigation while steering contained cohesive targets reactively through formation-level fields \cite{chipade2021multiagent}, reducing computational overhead but increasing susceptibility to local minima and deadlocks.

Purely reactive approaches also typically assume target cohesion while employing uniform strategies for both herder navigation and target steering. Pierson and Schwager \cite{piersonControllingNoncooperativeHerds2018} generate local viewpoints around obstacles to guide cohesive groups via unicycle dynamics, while Zhang et al. \cite{zhang2024distributed} apply identical radial potential fields to both herders and cohesive targets. These uniform approaches offer computational simplicity but may lack the specialized control needed for the distinct behavioral requirements of herders versus targets.

Learning-based approaches such as that of Zhi and Lien \cite{zhi2021learning} similarly focus on cohesive target groups, employing reinforcement learning where a single herder steers the collective while avoiding obstacles. Although effective in complex scenarios, learned policies lack interpretability and may not generalize beyond training conditions.

In contrast to these methods, our strategy addresses the more challenging case of non-cohesive targets while incorporating obstacle avoidance capabilities. 
It combines computational efficiency with enhanced robustness by employing differentiated, reactive control strategies for herders and targets. By avoiding map-based global planning while incorporating hybrid normal-tangential obstacle avoidance, we achieve low computational overhead while reducing the deadlock susceptibility that typically affects purely repulsive potential field methods.

\section{Preliminaries and problem statement}
\label{sec: mathematization}
\paragraph{Notation} For any two points $\mathbf{q},\mathbf{p} \in \mathbb{R}^2$, we denote their relative position vector as $\mathbf{d}_{\mathbf{q}\mathbf{p}}=\mathbf{q}-\mathbf{p}$.
Given a closed set $B \subset \mathbb{R}^2$ with boundary $\partial B$ and any point $\mathbf{q} \in \mathbb{R}^2$, the orthogonal projection of $\mathbf{q}$ onto $\partial B$ is defined as $\operatorname{proj}_{B}(\mathbf{q})
= \arg\min_{\mathbf{p}\in\partial B} \|\mathbf{q}-\mathbf{p} \|$. Given two closed sets $A, B \subset \mathbb{R}^2$ with boundaries $\partial A$ and $\partial B$ respectively, the distance between these two sets is defined as $ d(\partial A, \partial B)
= \inf_{\mathbf{x} \in \partial A, \mathbf{y} \in \partial B}
\|\mathbf{x} - \mathbf{y}\| $. 
We use $\mathrm{R}(\theta)$ to denote the standard planar rotation matrix by angle $\theta$.
When the subscript $\perp$ is used, $\mathrm{R}_{\perp}\theta)$ denotes the fixed rotation matrix
corresponding to a $\theta = \pm\frac{\pi}{2}$~rad rotation (perpendicular direction). 

\paragraph{Problem statement}
We consider a spatial domain $\Omega \equiv \mathbb{R}^2$, populated by $L$ obstacles, $N$ active herder agents and $M$ target agents, denoted by the sets $\mathcal{O}$, $\mathcal{H}$ and $\mathcal{T}$, respectively. We denote by $\mathbf{T}_a$, $a=1,\ldots, M$, the cartesian coordinates of the target agents, by $\mathbf{H}_i$, $i=1,\ldots,N$, those of the herders and by $\mathbf{P}_j$, $j=1,\dots,L$, those of the obstacles' centroid.
Obstacles are assumed to be convex, static in position, with boundaries such that  $ d(\partial O_i, \partial O_j) > \lambda_o + \epsilon_o$ $\forall i,j \in \mathcal{O}, i\neq j$, where $\lambda_o$ and $\epsilon_o$ are two positive constants. This separation constraint ensures that the influence regions of different obstacles do not overlap, simplifying the navigation dynamics and preventing conflicting repulsive forces. 
The goal is for the herders to corral and confine the targets within a desired region assumed to be, without lack of generality, a disc centered at the origin of radius $\rho_g$ \cite{aulettaHerdingStochasticAutonomous2022}. 

The target agents exhibit autonomous stochastic behavior without inherent coordination or communication capabilities.
Each target $T_a \in \mathcal{T}$ is influenced by any herder only within a distance $\lambda$, forming the neighbor set defined as:
\begin{equation}
    \mathcal{N}_H(T_a) = \{ H_i \in \mathcal{H} : \| \mathbf{H}_i - \mathbf{T}_a \| < \lambda \}.
\end{equation}
The dynamics of each target follows the model in \cite{lamaShepherdingControlHerdability2024}, with an additional term describing the local repulsion from nearby obstacles:
\begin{equation}
   \dot{\mathbf{T}}_a = \sqrt{2D }\boldsymbol{\mathcal{W}} 
   + \beta \sum_{i \in {N}_H(T_a)} \left( \lambda - \|\mathbf{d}_{\mathbf{H}_i\mathbf{T}_a}\| \right) 
   \hat{\mathbf{d}}_{\mathbf{H}_i\mathbf{T}_a}+ 
   \mathbf{F}^T_{a,O}
   \label{eq: target_dynamics}
\end{equation}
\noindent where ${\mathcal{W}}$  is a Brownian noise term with diffusion $D>0$, $\beta>0$ scales the repulsive drift induced by nearby herders, and $\mathbf{F}^T_{a,O}$ is the repulsive force felt by the targets near obstacles (see Fig. \ref{fig:obstacles_scheme} for a schematic representation of the forces at play).

Specifically, we set
\begin{equation}
 \mathbf{F}^T_{a,O}= - \sum_{j=1}^{L} \nabla U_j(\mathbf{T}_a)
\end{equation}
where $U_j$ is the repulsive potential field associated with obstacle $O_j \in \mathcal{O}$.

Following \cite{khatib1986real}, the repulsive potential associated with each obstacle $O_j$ is defined, for any point $\mathbf{q}\in \mathbb{R}^2$, as
\begin{equation}
U_j(\mathbf{q}) =
\begin{cases}
     \tfrac{k_o}{2} \left( \dfrac{1}{\|\mathbf{s}_j(\mathbf{q}) \|} - \dfrac{1}{\lambda_o} \right)^2, & \text{if } \|\mathbf{s}_j(\mathbf{q}) \| \leq \lambda_o, \\[1ex]
     0, & \text{if } \|\mathbf{s}_j(\mathbf{q}) \| > \lambda_o,
\end{cases}
\label{eq: obstacle_avoidance}
\end{equation}
where $\mathbf{s}_j(\mathbf{q}) = \mathbf{q} - \operatorname{proj}_{O_j}(\mathbf{q})$ is the vector from $\mathbf{q}$ to its orthogonal projection on obstacle $O_j$'s perimeter, and $\lambda_o>0$ denotes the obstacle influence radius. This potential is continuously differentiable for $\|\mathbf{s}_j(\mathbf{q}) \|>0$, diverges near the boundary of $O_j$, and vanishes outside the influence radius $\lambda_o$, thus preventing collisions while keeping the effect local.

We assume that each herder \(H_i \in \mathcal{H}\) is capable of sensing all targets as well as the other herders in the environment. Furthermore, when operating within the influence zone of an obstacle, a herder is assumed to have knowledge of the obstacle’s centroid and to be able to associate each point on the obstacle boundary with its corresponding centroid.

We assume the herders have a maximum speed $v_H>0$, with dynamics given by
\begin{equation}
\label{eq:herders_dynamics}
\mathbf{\dot{H}}_i=\mathbf{u}_i
\end{equation}
where the control input $\mathbf{u}_i$ must be designed so as to successfully drive targets towards the goal region in the presence of obstacles.

\begin{remark}
Note that as often done in the literature \cite{sebastian_adaptive_2022, zhang2024distributed, aulettaHerdingStochasticAutonomous2022}, we assume targets follow first-order dynamics under the kinematic assumption that their motion is primarily governed by instantaneous behavioral responses rather than inertial effects, which is reasonable for lightweight agents or when response times are much slower than the mechanical time constants \cite{albi2016}. The same assumption is also made for herders under the hypothesis that higher-level motion planning generates velocity commands that are executed by lower-level dynamic controllers \cite{olfatiFlocking2006}.
\end{remark}

\section{Obstacle-aware Shepherding strategy}
\label{sec:shep_obstacle_strategy}
When steering targets, herders must explicitly account for obstacle presence. We extend the target selection strategy from \cite{lamaShepherdingControlHerdability2024} to handle obstacle-rich environments through a two-stage process: target selection followed by obstacle-aware steering.
Each herder employs the same target selection rule as in \cite{lamaShepherdingControlHerdability2024}: it selects the farthest target from the goal, provided that target is not closer to any other herder in the environment. When no such target exists, the herder returns to the goal region to maintain containment while simply avoiding obstacles. The key extension lies in the steering phase, where herders must guide selected targets along feasible paths that circumnavigate obstacles when direct routes are blocked by positioning themselves in appropriate steering points. 

To achieve this enhanced behavior, we choose $\mathbf{u}_i$ in \eqref{eq:herders_dynamics} as
\begin{equation}
\label{eq:u_i}
    \mathbf{u}_i = (1 - \eta_i)\,\mathbf{F}_i(\mathbf{H}_i) 
    + \eta_i\, \mathbf{I}'_i(\mathbf{T}, \mathbf{H}, \mathcal{O})+ \mathbf{F}^H_{i,O}
\end{equation}
where $\eta_i \in \{0,1\}$ indicates whether $H_i$ is chasing a target ($\eta_i=1$) or not ($\eta_i=0$). (See Fig. \ref{fig:obstacles_scheme} for a schematic representation of the forces acting on the herders.)

If no target is selected ($\eta_i=0$), the herder returns to the goal region at a constant speed $v_H$ until reaching its boundary. This is achieved by setting:
\begin{equation}
    \mathbf{F}_i(\mathbf{H}_i) = 
    \begin{cases}
        -v_H \mathbf{H}_i, & \|\mathbf{H}_i\| > \rho_g, \\[3pt]
        0, & \|\mathbf{H}_i\| \leq \rho_g.
    \end{cases}
\end{equation}
Thus, idling herders do not wander but regroup near the goal region to keep the targets already driven therein confined.

\begin{figure}
    \centering
    \vspace{0.2cm}
    \includegraphics[width=0.9\linewidth]{figures/scheme_updated.pdf}
    \caption{Schematic representation of the forces acting on targets and herders according to \eqref{eq: target_dynamics} and \eqref{eq:herders_dynamics}-\eqref{eq:u_i} when moving in the environment, possibly interacting with each other and an obstacle. Herder $H_1$ is subject to the force $\mathbf{F}_1$ pulling it towards the goal region (in green) while being repelled by the obstacle (in red) through the force $\mathbf{F}_{1,O}^H$. Herder $H_2$ is moving towards the target $T_1$ in the absence of any nearby obstacle. Herder $H_3$ is moving towards target $T_2$ while avoiding the obstacle, placing itself at a tangential direction so as to make the target circumnavigate the obstacle. Herder $H_4$ is moving towards the goal region in the absence of nearby targets or obstacles.}
    \label{fig:obstacles_scheme}
\end{figure}

At the same time while navigating the environment herders need to avoid obstacles. This is achieved by setting  $\mathbf{F}^H_{i,O}$ in \eqref{eq:u_i} as a hybrid force combining normal and tangential components via a tunable parameter $\gamma \in [0,1]$:
\begin{equation}
    \mathbf{F}^H_{i,O}=
- \sum_{j=1}^{L} \Big( \gamma \nabla U_j(\mathbf{H}_i) + (1 - \gamma) R_\perp(\theta_i)\,\nabla U_j(\mathbf{H}_i) \Big)
\end{equation}

The first term provides normal repulsion using the same potential field form adopted for the targets in \eqref{eq: obstacle_avoidance}, while the second term represents an additional tangential force obtained by rotating the normal component by an angle $\theta_i$, chosen as
\begin{equation}
    \theta_i = 
    \begin{cases}
        \tfrac{\pi}{2}, & \mathbf{d}_{\mathbf{H}_i \mathbf{P}_j} \times \mathbf{d}_{\mathbf{C}_i \mathbf{P}_j} > 0, \\[6pt]
        -\tfrac{\pi}{2}, & \mathbf{d}_{\mathbf{H}_i \mathbf{P}_j} \times \mathbf{d}_{\mathbf{C}_i \mathbf{P}_j} \leq 0,
    \end{cases}
\end{equation}
where $\mathbf{C}_i$ denotes the herder's steering point relative to its selected target if it is chasing a target, and $\mathbf{C}_i = \mathbf{0}$ otherwise.

The tangential component prevents deadlock situations that can occur with purely repulsive normal forces, enabling herders to slide along obstacle boundaries toward their objectives. The rotation direction is determined by the relative geometry to avoid unnecessary detours and prevent oscillations.

If a target $T_i^*$ is selected ($\eta_i=1$), while still avoiding obstacles through the force $\mathbf{F}^H_{i,O}$, herders also need to drive the chosen target towards the goal region. This is achieved by activating the second term on the right-hand side of \eqref{eq:u_i} which we set as:
\begin{equation}
\mathbf{I}'_i(\mathbf{T}, \mathbf{H}, \mathcal{O}) 
= - \alpha \left( \mathbf{H}_i - \Big[ \mathbf{T}_i^* + \delta \big( \mu_i \hat{\mathbf{T}}_i^* + (1-\mu_i) \hat{\boldsymbol{\nu}}_i \big) \Big] \right).
\end{equation}
Here, \(\alpha > 0\) denotes a control gain, \(\delta < \lambda\) specifies the offset distance, \(\hat{\mathbf{T}}_i^*\) is the unit vector pointing from the goal center toward the target, and \(\hat{\boldsymbol{\nu}}_i\) represents the tangent direction along the obstacle boundary.

The binary variable $\mu_i$ is set to $\mu_i=1$ for a purely straight push in the absence of nearby obstacles, and to $\mu_i=0$ when the target is at a distance $\lambda_{o}+\epsilon_o$ from an obstacle blocking its path to the goal. When this happens, the herder need to push the target tangentially to the obstacle, allowing the target to avoid it while being driven towards the goal region.

The boundary-tangent direction is defined as a rotation of the vector 
$\mathbf{s}_{\hat{j}}(\mathbf{T}_i^*) = \mathbf{T}_i^* - \operatorname{proj}_{O_{\hat{j}}}(\mathbf{T}_i^*)$,
obtained by setting:
\begin{equation}
    \boldsymbol{\nu}_i =
    \begin{cases}
        R_\perp\!\left( -\tfrac{\pi}{2} \right)\,\mathbf{s}_{\hat{j}}(\mathbf{T}_i^*), & 
        \mathbf{d}_{\mathbf{T}_i^* \mathbf{P}_{\hat{j}}} \times \mathbf{d}_{\mathbf{O}\mathbf{P}_{\hat{j}}} > 0, \\[1ex]
        R_\perp\!\left( \tfrac{\pi}{2} \right)\,\mathbf{s}_{\hat{j}}(\mathbf{T}_i^*), & 
        \mathbf{d}_{\mathbf{T}_i^* \mathbf{P}_{\hat{j}}} \times \mathbf{d}_{\mathbf{O}\mathbf{P}_{\hat{j}}} \leq 0,
    \end{cases}
\end{equation}
so that the target is consistently driven along the most favorable direction.

\section{Numerical validation}

\begin{figure}
    \centering
    \vspace{0.2cm}
    \includegraphics[width=0.8\linewidth]{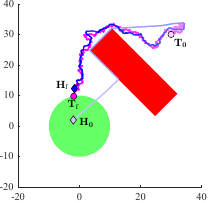}
    \caption{Example of obstacle-avoidance strategy in a single herder and single target setup. In order to drive the target, the herder first circumnavigates the obstacle, then it pushes the target tangentially. 
    Color gradients represent the progression of positions over time, going from $t=0$ to $t=t_f$.}
    \label{fig:traj_single_matlab}
\end{figure}

Numerical experiments are conducted in MATLAB to evaluate the proposed obstacle-aware shepherding strategy.

The simulation environment consists of a circular domain 
\(\Omega_0 \subset \Omega \) with radius \(\rho_0 = 50\,\text{m}\), from which initial positions of herders and targets are uniformly sampled. A single rectangular obstacle with dimensions 
\(30\,\text{m} \times 10\,\text{m}\), centered at \((15\,\text{m}, 15\,\text{m})\), with its major axis rotated of an angle of \( 3\pi/4\) radians from the positive horizontal axis to create a challenging navigation scenario that blocks direct paths to the goal region.

Figure \ref{fig:traj_single_matlab} illustrates a representative trajectory demonstrating the proposed obstacle circumnavigation strategy. The herder initially approaches the obstacle and utilizes the tangential force component to navigate around its boundary, maintaining forward progress toward the target. Once positioned appropriately, the herder applies repulsive forces to guide the target along a similar circumnavigation path, successfully steering it into the goal region while avoiding collision with the obstacle.

\begin{figure}
    \centering
    \vspace{0.2cm}
    \subfloat[]
{
    \includegraphics{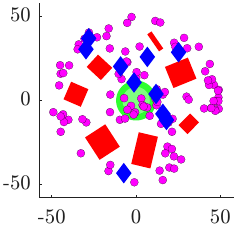}
    \label{fig:initial_multiobstacle}
}
    \subfloat[]
{
    \includegraphics{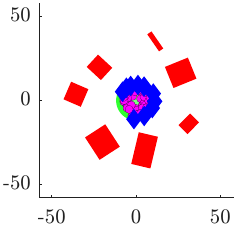}
    \label{fig:final_multiobstacle}
}\\
    \subfloat[]
{
    \includegraphics{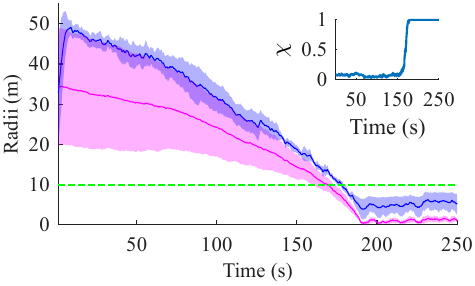}
    \label{fig:radii_matlab_multi}
}
\caption{Validation of the proposed shepherding strategy in a scenario with 
\(\numherders = 10\) herders, \(\numtargets = 100\) targets and \(L = 7\) rectangular obstacles with width and height uniformly sampled from \([8,20]\)\,m and \([2,20]\)\,m, respectively. (a) Initial positions of herders (blue diamonds) and targets (magenta points). (b) Configuration at time \( 177.71\)\,s, corresponding to the first time instant at which all targets enter the goal region. 
(c) Temporal evolution of the mean distance from the goal region center for herders (blue) and targets (magenta), with the corresponding standard deviations (shaded areas), relative to the goal region radius (green dashed line). 
The inset shows the evolution of the fraction of captured targets \(\chi\), which 
successfully converges to unity.
}
\end{figure}

The proposed strategy scales naturally to large multi-agent scenarios through decentralized decision-making, where each herder independently selects targets using simple local rules. Fig.~\ref{fig:initial_multiobstacle} shows the initial configuration of 10 herders and 100 targets in the presence of \(L = 7\) obstacles whose width and height are independently sampled from uniform distributions over the intervals \([8,20]\)\,m and \([2,20]\)\,m, respectively. Under the proposed control law, all targets first enter the goal region at time \(177.71\)\,s, as illustrated in Fig.~\ref{fig:final_multiobstacle}. The temporal evolution of the agents’ distances from the goal over the full simulation horizon is reported in Fig.~\ref{fig:radii_matlab_multi}, confirming both convergence and subsequent containment within the goal region. The 10:100 herder-to-target ratio is based on the herdability analysis in~\cite{lamaShepherdingControlHerdability2024}, ensuring sufficient herding capacity for the obstacle-free baseline case.

To investigate the impact of non-cohesive target behavior, we compare our method against the state-of-the-art approach by Pierson and Schwager \cite{piersonControllingNoncooperativeHerds2018}. Their method assumes cohesive targets and employs unicycle dynamics, positioning herders along an arc centered at the targets' center of mass to push the group toward the goal. For obstacle avoidance, their approach generates waypoints to guide the unicycle around obstacles. Since the original work does not specify the waypoint generation algorithm, we implement the $A*$ path planning algorithm to ensure fair comparison through optimal path generation. 

Over 50 simulations with randomly initialized conditions in an environment containing a single rectangular obstacle, our strategy consistently succeeded in herding all targets, whereas, as expected, the approach of \cite{piersonControllingNoncooperativeHerds2018} failed consistently, capturing only $\chi = 0.05 \pm 0.07$ of the targets on average. This stark performance difference demonstrates that obstacle avoidance capabilities alone are insufficient for non-cohesive targets. In this latter case, effective shepherding requires individual target consideration rather than center-of-mass control strategies designed for cohesive groups. Model parameters and simulation details are provided in the Appendix.

\section{Experimental validation}
To evaluate the effectiveness of the proposed control strategy, we conducted real-world experiments using an heterogeneous group of mobile robots comprising \textit{TurtleBot4} as herders and \textit{Osoyoo Robotic Car V4.0} as targets.

The \textit{TurtleBot4}, equipped with wheel encoders and factory-calibrated odometry, provides precise motion control suitable for executing the deterministic herder control policy. The \textit{Osoyoo Robotic Car V4.0} is a low-cost differential-drive platform based on a Raspberry Pi that lacks wheel encoders, resulting in less precise motion control. This inherent imprecision makes it an appropriate platform for representing the stochastic, non-cohesive target behavior modeled in our approach.

Experiments were carried out in a $2\, \text{m} \times 2\, \text{m}$ indoor arena containing a rectangular obstacle of size $0.7\, \text{m} \times 0.25\, \text{m}$ and a goal region of radius $0.35\, \text{m}$. The setup was monitored by a motion-tracking system providing ground-truth poses of all robots and obstacles. These data were processed by a central computer, resulting in a pseudo-distributed architecture: although computations were centralized, each robot acted solely on local information.  

\subsection{Model adaptation for practical implementation}
The theoretical model assumes point agents governed by single-integrator dynamics. In practice, robots have finite dimensions requiring model adaptation to account for physical embodiment. Without this adaptation, agents would attempt to occupy the same spatial location, leading to unrealistic overlapping behaviors.
To prevent inter-agent collisions, we introduce pairwise repulsive forces between agents of the same type. Following the obstacle avoidance potential field design, this repulsive interaction vanishes when the inter-agent distance $\| \mathbf{d}_{\mathbf{q}_i \mathbf{q}j}\|$ exceeds threshold $d_{th}$ and diverges as agents approach each other. For two agents positioned at  $\mathbf{q}_i$ and $\mathbf{q}_j$, the repulsive potential is:
\begin{equation}
V_{ij} =
\begin{cases}
     \tfrac{k_d}{2} \left( \dfrac{1}{\| \mathbf{d}_{\mathbf{q}_i \mathbf{q}_j}\|} - \dfrac{1}{d_{th}} \right)^2, & \text{if } \| \mathbf{d}_{\mathbf{q}_i \mathbf{q}_j}\| \leq d_{th}, \\[1ex]
     0, & \text{if } \| \mathbf{d}_{\mathbf{q}_i \mathbf{q}_j}\| > d_{th},
\end{cases}
\end{equation}
with the corresponding repulsive force $\mathbf{G}^T_{ij} = -\nabla V_{ij}$ if $i$ and $j$ are targets,  $\mathbf{G}^H_{ij}$ otherwise.

Consider herder pursuing its selected target $T_i^*$, with $\mathbf{C}_i$ denoting the steering point relative to $T_i^*$. To prevent collision during approach while maintaining effective control, an orbiting term is introduced that enables the herder to circulate around the target when a direct approach is not feasible.

\begin{figure}
    \centering
    \vspace{0.2cm}
    \subfloat[]
{
    \includegraphics[width=0.45\linewidth]{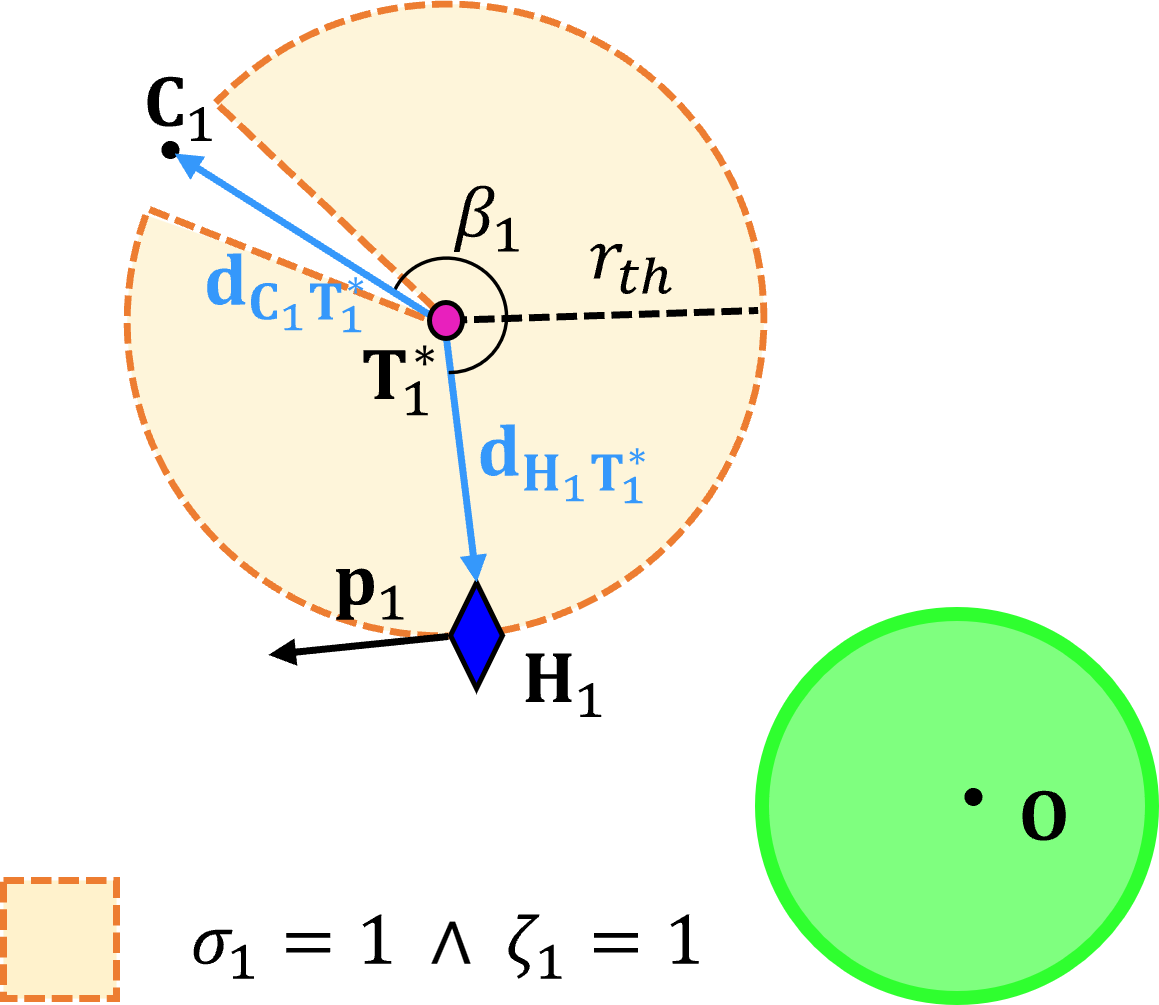}
    \label{fig:orbiting_logic}
}
    \subfloat[]
{
    \includegraphics[width=0.45\linewidth]{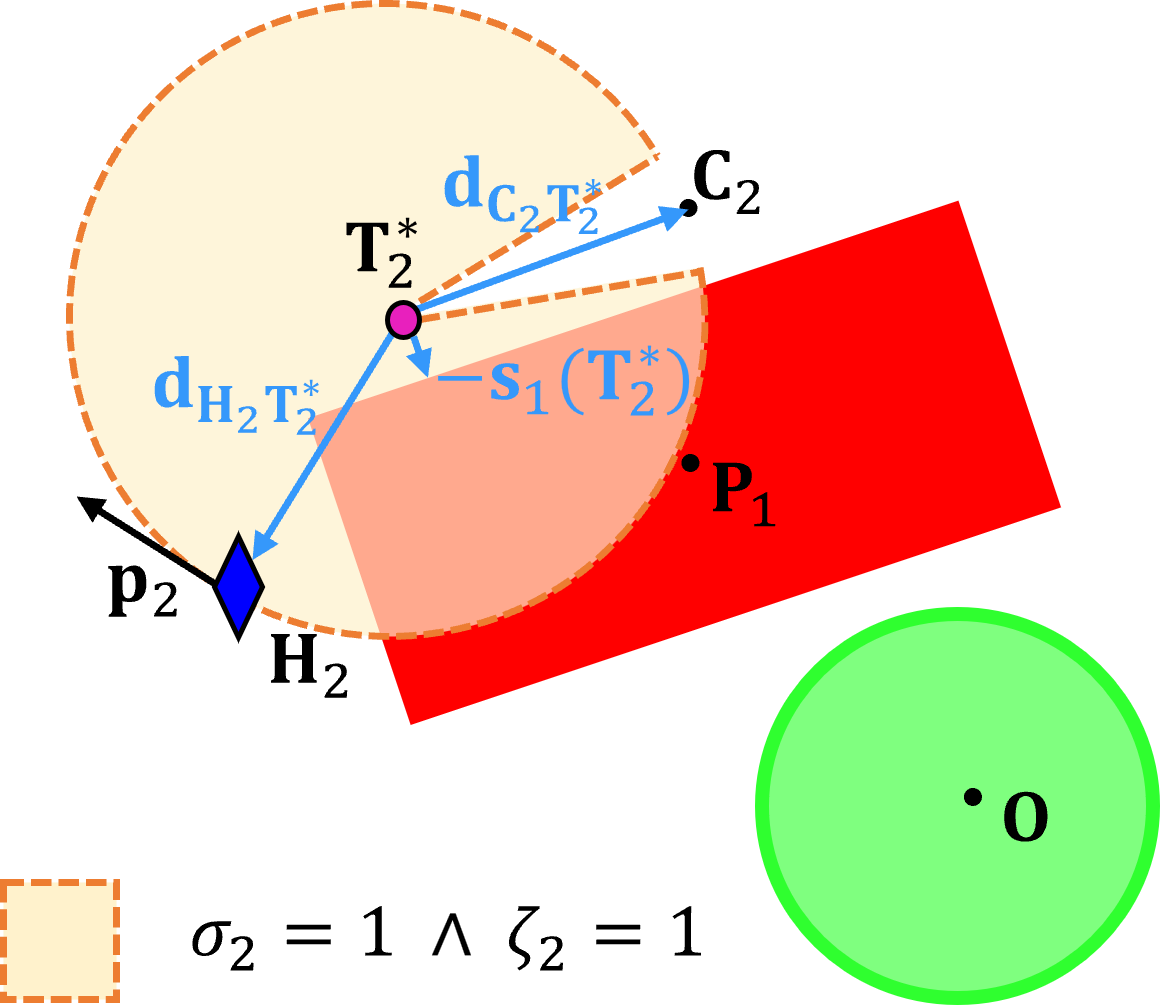}
    \label{fig:orbiting_logic_obs}
}
\caption{Orbiting direction selection for herder $H_i$ pursuing its assigned target $T_i^*$. 
    (a) Obstacle-free case: the tangential orbiting direction is determined by the relative orientation between $\mathbf{d}_{\mathbf{H}_i \mathbf{T}_i^*}$ and $\mathbf{d}_{\mathbf{C}_i \mathbf{T}_i^*}$, yielding $\phi_i = \pm \frac{\pi}{2}$ according to Eq.~\eqref{eq: phi_i} and enabling circulation toward the desired offset position $\mathbf{C}_i$. 
    (b) Target in proximity of obstacle $O_j$: the nominal orbiting direction may steer the herder toward collision-prone configurations due to the obstacle-induced constraint represented by $-\mathbf{s}_j(\mathbf{T}_i^*)$. 
    In this case, the modulation term $\psi_i$, expressed in Eq.~\eqref{eq: psi_i}, reverses the orbiting direction to ensure safe circulation of $T_i^*$ around the obstacle. 
    The shaded regions denote configurations in which the tangential contribution in $\mathbf p_i$ dominates the radial one, corresponding to the simultaneous saturation of $\sigma_i$ in Eq.~\eqref{eq: sigma_i} and $\zeta_i$ in Eq.~\eqref{eq: zeta_i}.}
    \label{fig:orbiting}

\end{figure}

In the absence of nearby obstacles, the orbiting direction is determined by $\mathbf{p}_i = \alpha_o \mathrm{R}_\perp(\phi_i) \hat{\mathbf{d}}_{\mathbf{H}_i \mathbf{T}i^*} + \alpha_r  \hat{\mathbf{d}}_{\mathbf{H}_i \mathbf{T}i^*}\left( 1- {\| \mathbf{d}_{\mathbf{H}_i \mathbf{T}i^*}\|}/{r_{th}} \right)$, where $r_{th}$ is a distance threshold,  $\alpha_o>0$ is the orbiting tangential gain, $\alpha_r$ is a radial gain regulating $\| \mathbf{d}_{\mathbf{H}_i \mathbf{T}i^*}\|$ to $r_{th}$, and the rotation angle is given by:
\begin{equation}
\phi_i = 
    \begin{cases}
        \frac{\pi}{2}, & \text{if } \mathbf{d}_{\mathbf{H}_i \mathbf{T}_i^*} \times \mathbf{d}_{\mathbf{C}_i \mathbf{T}_i^*} \geq 0\\
        -\frac{\pi}{2}, & \text{if } \mathbf{d}_{\mathbf{H}_i \mathbf{T}_i^*} \times \mathbf{d}_{\mathbf{C}_i \mathbf{T}_i^*} <  0\\
    \end{cases}
    \label{eq: phi_i}
\end{equation}
The resulting orbiting direction is illustrated in Fig.~\ref{fig:orbiting_logic} for the obstacle-free case.
However, when the selected target $T_i^*$ is in proximity of an obstacle, the nominal orbiting direction may induce configurations that lead to collisions during circulation. This situation is depicted in Fig.~\ref{fig:orbiting_logic_obs}. To address this issue, the rotation angle is modified as $\tilde{\phi}_i = \psi_i \phi_i$, where $\psi_i$ is defined as follows:
\begin{equation}
\psi_i =
\begin{cases}
-1, &
\begin{aligned}[t]
&\bigl|\angle(\mathbf d_{\mathbf H_i\mathbf T_i^*}, \mathbf d_{\mathbf C_i\mathbf T_i^*})\bigr| > \tfrac{\pi}{2}\\
&\land\ \bigl| \angle(-\mathbf{s}_j(\mathbf T_i^*), \mathbf d_{\mathbf{H}_i\mathbf{T}_i^*})\bigl| < \tfrac{\pi}{2}
\end{aligned}
\\[1ex]
+1, & \text{otherwise.}
\end{cases}
\label{eq: psi_i}
\end{equation}
which allows the herder to circulate the target without colliding with obstacles. 

The orbiting term is modulated by two scalar weights, denoted by $\sigma_i, \zeta_i \in [0,1]$, that activate the orbiting action only when needed: the first depends on the distance between the herder and its selected target, and the second on the alignment between the vectors $\mathbf{d}_{\mathbf{H}_i \mathbf{T}^*_i}$ and $\mathbf{d}_{\mathbf{C}_i \mathbf{T}^*_i}$.

Specifically, let $r_{th}>0$ be a distance threshold and $\epsilon_h>0$ a buffer distance, $\sigma_i$ is given by:
\begin{equation}
    \sigma_i = \begin{cases}
        0, & \text{if }\|\mathbf{d}_{\mathbf{H}_i \mathbf{T}_i^*} \| \geq r_{th} + \epsilon_h \\
        1, & \text{if } \|\mathbf{d}_{\mathbf{H}_i \mathbf{T}_i^*} \| \leq r_{th} \\
        \frac{\|\mathbf{d}_{\mathbf{H}_i \mathbf{T}_i^*} \| -r_{th}-\epsilon_h}{\epsilon_h}, & \text{if } r_{th}< \|\mathbf{d}_{\mathbf{H}_i \mathbf{T}_i^*} \| < r_{th} + \epsilon_h
        
    \end{cases}
    \label{eq: sigma_i}
\end{equation}
Similarly, given $\beta_i = \angle \left( \mathbf{d}_{\mathbf{H}_i \mathbf{T}_i^*}, \mathbf{d}_{\mathbf{C}_i \mathbf{T}_i^*}\right)$ and two angle thresholds $\beta_{orb}$ and $\beta_{th}$ such that  $0<\beta_{orb}<\beta_{th}$, $\zeta_i$ is defined as: 
\begin{equation}
    \zeta_i = \begin{cases}
        0, & \text{if } |\beta_i| \leq \beta_{orb}\\
        1, & \text{if }|\beta_i| \geq \beta_{th} \\
        \frac{|\beta_i|- \beta_{orb}}{\beta_{th}-\beta_{orb}}, & \text{if } \beta_{orb} < |\beta_i|<\beta_{th}
    \end{cases}
    \label{eq: zeta_i}
\end{equation}
The region corresponding to the simultaneous saturation of both $\sigma_i$ and $\zeta_i$ (i.e., $\sigma_i = \zeta_i = 1$) is highlighted in yellow in Fig.~\ref{fig:orbiting}.

Given the additional orbiting term, $d(\partial O_i, \partial O_j)$ $\forall i,j \in O, i\neq j$, has to be increased by $d_{th}$ to enable target circulation. 

Therefore, for experimental implementation, the control law in Eq.~\eqref{eq:u_i} is modified to incorporate the physical robot constraints and orbiting behavior described above:
\begin{align}
\label{eq:u_i_tilde}
\tilde{\mathbf{u}}_i 
    &= (1 - \eta_i)\,\mathbf{F}_i(\mathbf{H}_i) 
     + (1-\sigma_i) \,\eta_i\, \mathbf{I}'_i(\mathbf{T}, \mathbf{H}, \mathcal{O}) \notag \\
    &\quad + \mathbf{F}^H_{i,O} 
     + \sum_{j\in \mathcal{H}} \mathbf{G}^H_{ij} 
     + \sigma_i \, \zeta_i \,\mathbf{p}_i
\end{align} 
\noindent where the function $\mathbf{I}_i'$ is also blended with $\sigma_i$ to prevent overlapping. 

In addition, given the inaccurate motion of the \textit{Osoyoo} robots,  $D$ in Eq.~\eqref{eq: target_dynamics} is set to 0 so that no additional noise is artificially introduced, and inter-agent repulsive forces are added for practical implementation. The target dynamics becomes:
\begin{equation}
   \dot{\tilde{\mathbf{T}}}_a = \beta \sum_{i \in {N}_H(T_a)} \left( \lambda - \|\mathbf{d}_{\mathbf{H}_i\mathbf{T}_a}\| \right) 
   \hat{\mathbf{d}}_{\mathbf{H}_i\mathbf{T}_a}+ 
   \mathbf{F}^T_{a,O}+  \sum_{j\in \mathcal{T}} \mathbf{G}^T_{aj}
   \label{eq: target_dynamics2}
\end{equation}

Once the single-integrator velocities are computed, a unicycle mapping has to be applied to obtain feasible linear and angular commands \cite{Siciliano2009}.
Specifically, let $\mathbf{u}_{si} = [u_x \; u_y]^T$ be the control input generated by the single-integrator model for a robot located at position $\mathbf{p} = [x \; y]^T$ with orientation $\varphi$. 
The nominal linear and angular velocities $(v,\omega)$ are obtained via
\begin{equation}
\begin{bmatrix}
v_{nom} \\ \omega_{nom}
\end{bmatrix}
= \begin{bmatrix}
\cos\varphi & \sin\varphi \\
-\tfrac{\sin\varphi}{d} & \tfrac{\cos\varphi}{d}
\end{bmatrix}
\mathbf{u}_{si},
\label{eq:unicycle_mapping}
\end{equation}
\noindent where $d>0$ is a tuning parameter representing the look-ahead distance of the unicycle mapping. In addition, given the actuator limits of the robot wheels $|v_L|, |v_R| \le v_{\max}$ and the wheelbase $l$, a scaling factor $s$ is applied:
\begin{equation}
s \;=\; \min\!\left\{\,1,\;
\frac{v_{\max}}{\left|\,v_\mathrm{nom}+\frac{l}{2}\,\omega_\mathrm{nom}\,\right|+\varepsilon},\;
\frac{v_{\max}}{\left|\,v_\mathrm{nom}-\frac{l}{2}\,\omega_\mathrm{nom}\,\right|+\varepsilon}
\right\},
\label{eq:scaling_factor}
\end{equation}
\noindent with $\varepsilon>0$ arbitrarily small. This ensures that the commanded velocities respect the physical limits of both robot wheels 
while preserving the direction of the nominal control input.

\subsection{Gazebo simulation}
As an intermediate step between theoretical analysis and experimental implementation, the robotic system described above is simulated in a Gazebo/ROS-2 environment (whose details are reported in Appendix). While the \textit{TurtleBot4} model is publicly available, the \textit{Osoyoo} implementation required a system identification procedure from the real robot, in order to determine its dynamical parameters. 

The control strategy from Sec. \ref{sec:shep_obstacle_strategy}, developed on point-particles, is enhanced via the low-level controllers from the previous paragraph, to make it compliant with the differential-drive kinematics of the mobile robots.

\begin{figure}
    \centering
    \vspace{0.2cm}
    \subfloat[]
{
    \includegraphics[width=0.9\linewidth]{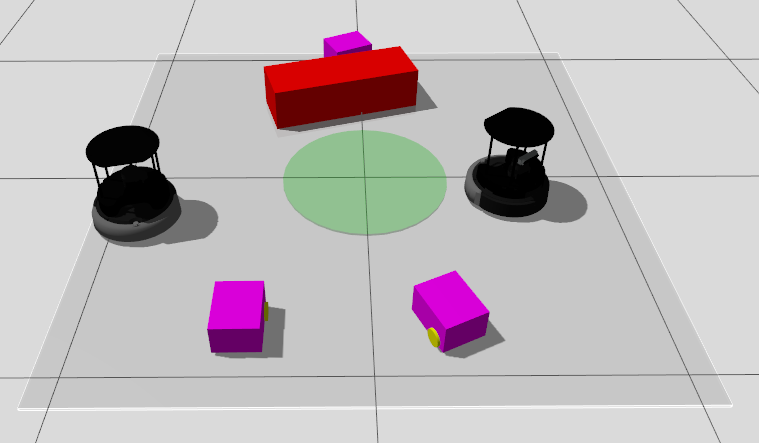}
    \label{fig:gazebo_snapshot}
}\\

    \subfloat[]
{
    \includegraphics[width=0.9\linewidth]{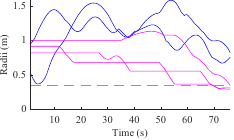}
    \label{fig:gazebo_radii}
}

    \caption{Gazebo simulation of the obstacle-aware shepherding strategy with two \textit{TurtleBot4} herders and three \textit{Osoyoo} targets, controlled via a ROS 2 control architecture. (a) Snapshot from Gazebo depicting initial robot configuration. (b) Time evolution of the distances of herders (blue) and targets (magenta) from the goal center. The green dashed line denotes the goal radius ($\goalradius = 0.35$~m).
}
    \label{fig:gazebo_sim}
\end{figure}

Fig. \ref{fig:gazebo_snapshot} shows the initial robots' configuation in a trial with two \textit{TurtleBot4} herders and three \textit{Osoyoo} targets, while Fig. \ref{fig:gazebo_radii} depicts the agents' distances from the goal center over an example run, showing how herders effectively confine the targets (see supplementary videos).

\subsection{Real robot experiments}
\label{sec: exp}
\begin{figure}
    \centering
    \vspace{0.2cm}
    \subfloat[]
{
    \includegraphics[width=0.4\linewidth]{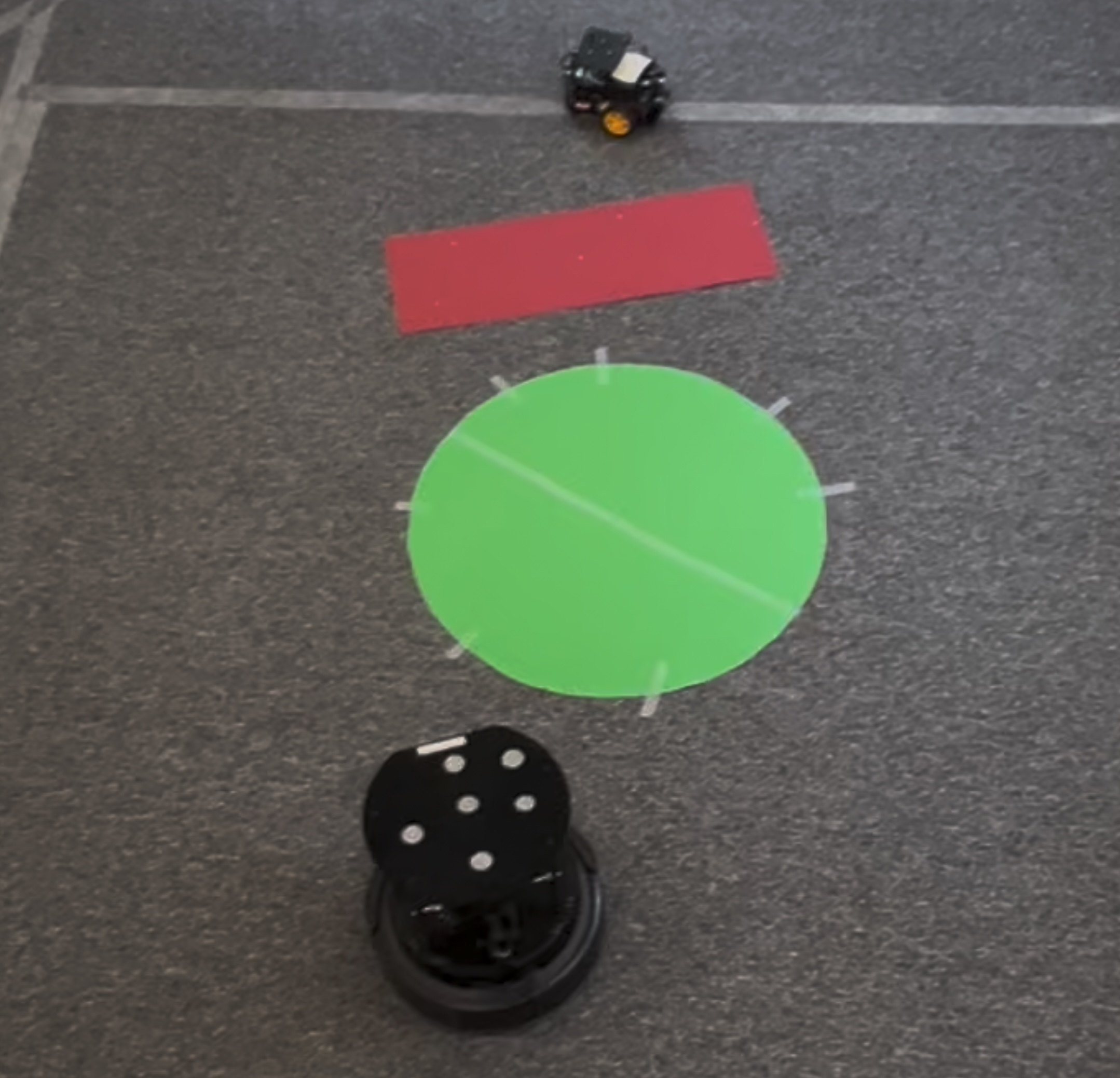}
    \label{fig:robot_initial_single}
}
    \subfloat[]
{
    \includegraphics[width=0.4\linewidth]{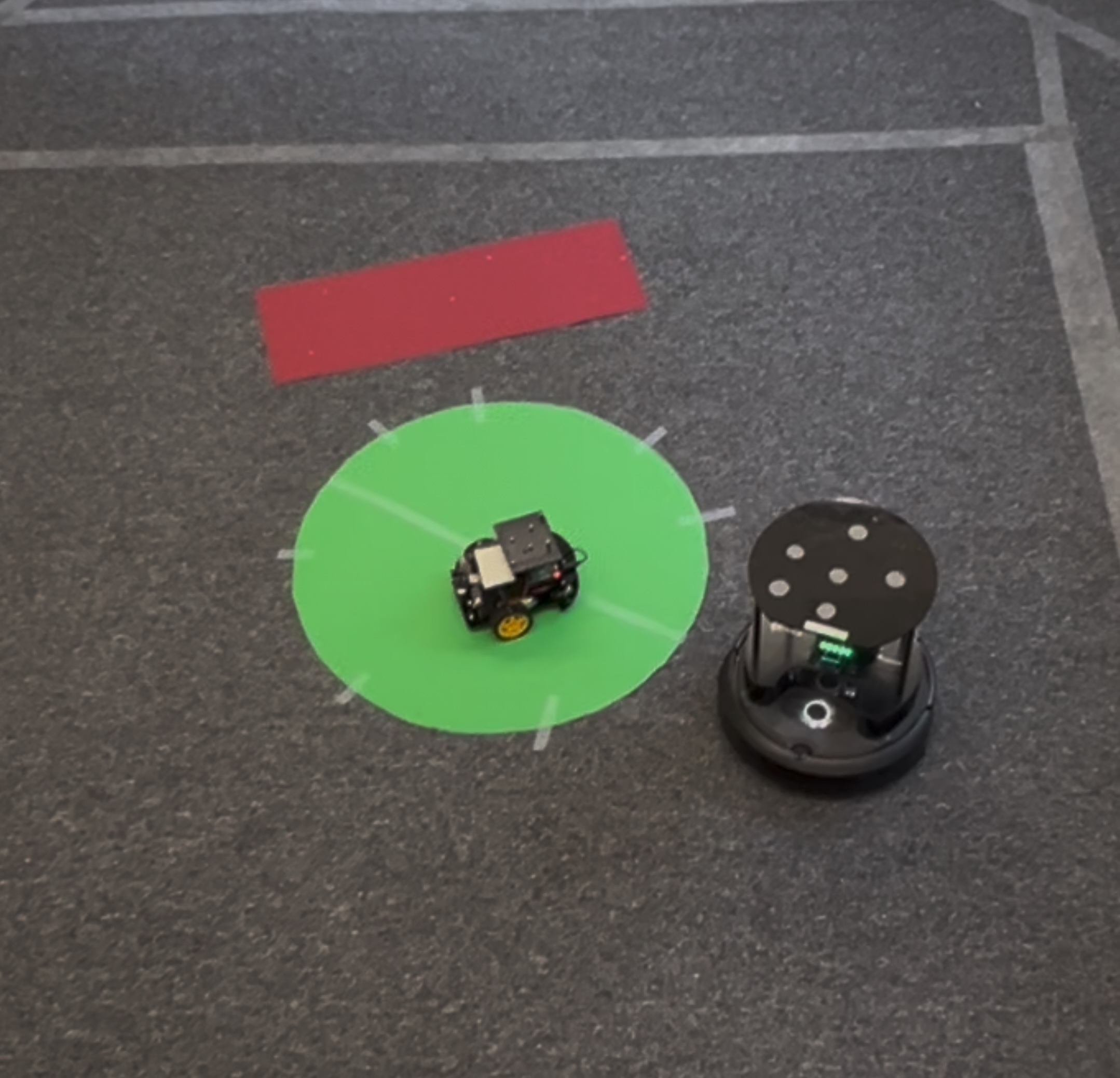}
    \label{fig:robot_final_single}
}\\
    \subfloat[]
{
    \includegraphics[width=0.8\linewidth]{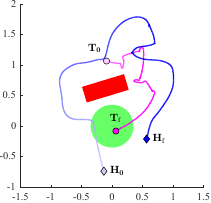}
    \label{fig:robot_traj}
}
\caption{
Experimental validation of the obstacle-aware shepherding strategy with one \textit{TurtleBot4} herder and one \textit{Osoyoo} target. (a)~Initial configuration with target positioned behind a rectangular obstacle. (b)~Final configuration showing the target successfully contained within the goal region. (c)~Robots' trajectories over time. Color gradients represent the progression of positions over time, going from $t=0$ -- agents indicated as $\mathbf{H}_0=\mathbf{H}(0)$ and $\mathbf{T}_0=\mathbf{T}(0)$ -- to $t=t_f$ -- agents indicated as $\mathbf{H}_f=\mathbf{H}(t_f)$ and $\mathbf{T}_f=\mathbf{T}(t_f)$.
}
    \label{fig:robot_single}
\end{figure}

Next, we validate the proposed strategy on the real robotic platform, using the set of parameters reported in the Appendix.  
In the first experiment, a single \textit{TurtleBot4} herder and one \textit{Osoyoo} target are deployed. The target is initially positioned behind the obstacle (Fig.~\ref{fig:robot_initial_single}), and the herder successfully reaches it, guiding it toward the goal region while avoiding collisions (Fig.~\ref{fig:robot_final_single}). Figure~\ref{fig:robot_traj} shows the trajectories of the robots: the herder’s path clearly illustrates how it circumvents the obstacle by moving around it, then pushes the target tangentially along the most favorable side of the obstacle, before ultimately driving it into the goal region (see supplementary videos).  

A second experiment considers a more challenging configuration with two herders and three targets, under the same obstacle placement. In this scenario, the herders autonomously select suitable targets and coordinate to effectively contain and steer the group toward the goal region.  

\begin{figure}
    \centering
    \vspace{0.2cm}
    \subfloat[]
{
    \includegraphics[width=0.4\linewidth]{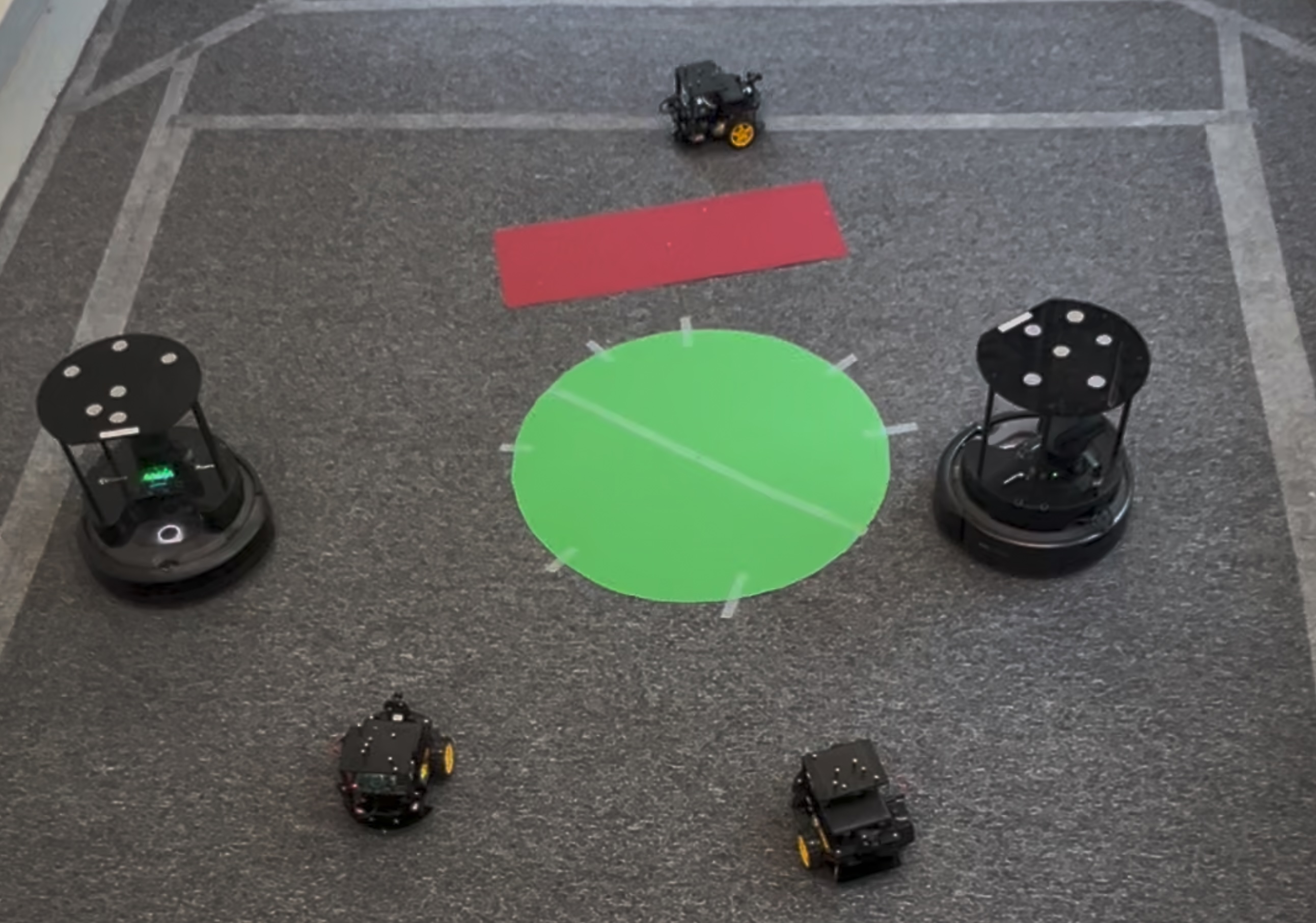}
    \label{fig:robot_initial_multi}
}
    \subfloat[]
{
    \includegraphics[width=0.4\linewidth]{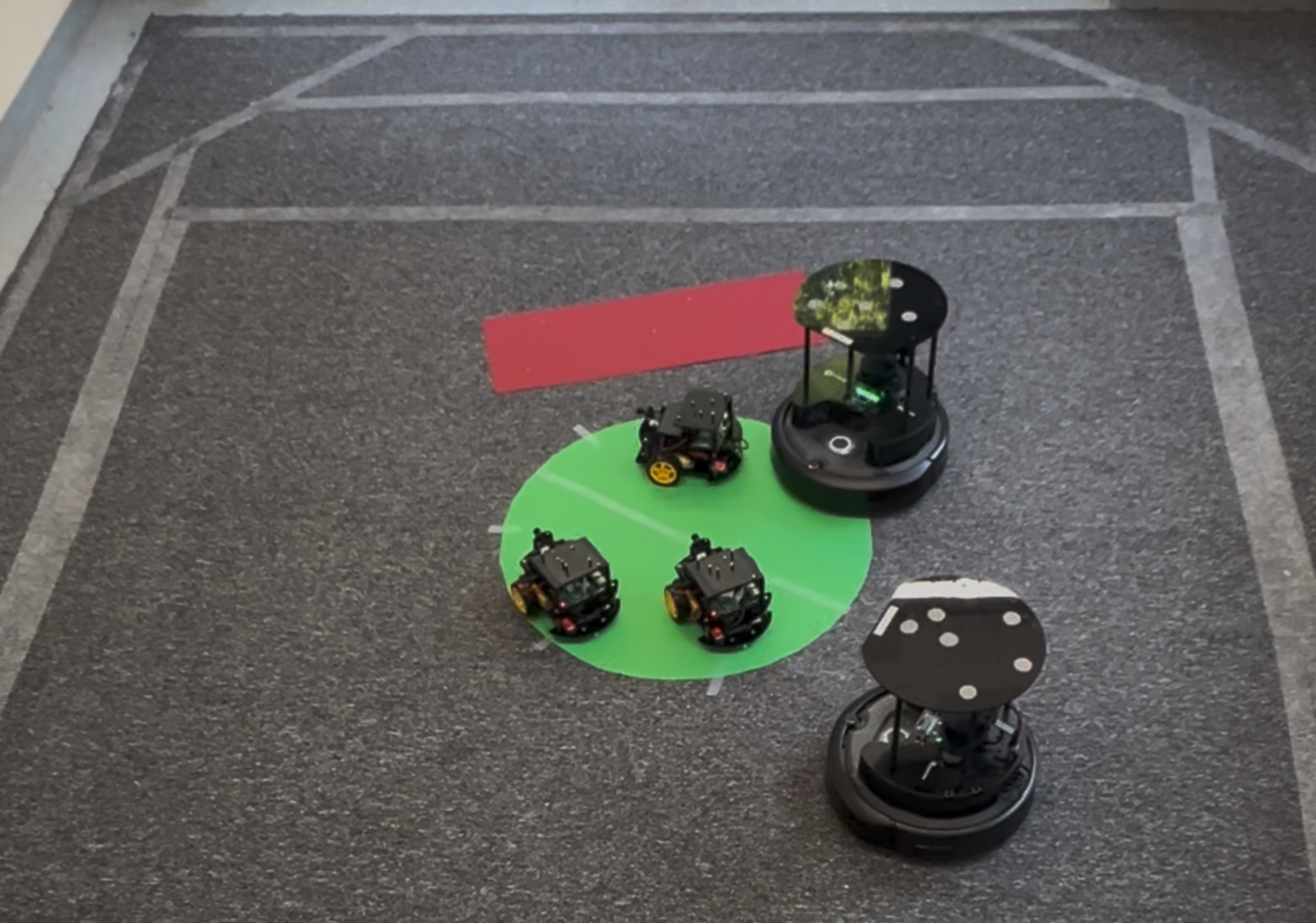}
    \label{fig:robot_final_multi}
}\\
    \subfloat[]
{
    \includegraphics[width=0.9\linewidth]{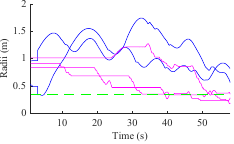}
    \label{fig:robot_radii}
}
\caption{
Experimental validation of the obstacle-aware shepherding strategy with one \textit{TurtleBot4} herder and one \textit{Osoyoo} target. (a) Initial configuration with a target robot placed behind an obstacle. (b) Final configuration showing all targets successfully contained within the goal region. (c) Time evolution of the radial distances of herders (blue) and targets (magenta) from the goal center. The green dashed line denotes the goal radius ($\goalradius = 0.35$~m).
}
    \label{fig:robot_multi}
\end{figure}

Figures~\ref{fig:robot_initial_multi}--\ref{fig:robot_final_multi} illustrate the initial and final configurations of the robots, while Fig.~\ref{fig:robot_radii} highlights the success of our strategy by depicting the agents' distances from the goal center in time (see supplementary videos).

\section{Conclusions}
This paper presents a control strategy for multi-agent shepherding of non-cohesive targets in obstacle-rich environments. The approach combines repulsive and tangential forces to enable herders to guide targets around obstacles through circumnavigation maneuvers, addressing limitations of existing methods that assume cohesive target behavior.
Numerical simulations demonstrate superior performance compared to state-of-the-art approaches, achieving higher target confinement rates with improved efficiency. Experimental validation using heterogeneous robot platforms confirms practical viability under real-world constraints including actuator limitations and sensing uncertainties.
Future work should address dynamic obstacles, theoretical convergence analysis, and extension to 3D environments.
\vspace*{0.15cm}\\
\textbf{Acknowledgments.} A.R.\,acknowledges support from Germany's DFG Excellence Strategy -- EXC 2117-422037984.

\addtolength{\textheight}{-2cm}   
\bibliographystyle{IEEEtran}

\begin{thebibliography}{10}
\providecommand{\url}[1]{#1}
\csname url@samestyle\endcsname
\providecommand{\newblock}{\relax}
\providecommand{\bibinfo}[2]{#2}
\providecommand{\BIBentrySTDinterwordspacing}{\spaceskip=0pt\relax}
\providecommand{\BIBentryALTinterwordstretchfactor}{4}
\providecommand{\BIBentryALTinterwordspacing}{\spaceskip=\fontdimen2\font plus
\BIBentryALTinterwordstretchfactor\fontdimen3\font minus \fontdimen4\font\relax}
\providecommand{\BIBforeignlanguage}[2]{{%
\expandafter\ifx\csname l@#1\endcsname\relax
\typeout{** WARNING: IEEEtran.bst: No hyphenation pattern has been}%
\typeout{** loaded for the language `#1'. Using the pattern for}%
\typeout{** the default language instead.}%
\else
\language=\csname l@#1\endcsname
\fi
#2}}
\providecommand{\BIBdecl}{\relax}
\BIBdecl

\bibitem{long2020comprehensive}
N.~K. Long, K.~Sammut, D.~Sgarioto, M.~Garratt, and H.~A. Abbass, ``A comprehensive review of shepherding as a bio-inspired swarm-robotics guidance approach,'' \emph{IEEE Transactions on Emerging Topics in Computational Intelligence}, vol.~4, no.~4, pp. 523--537, 2020.

\bibitem{strombomSolvingShepherdingProblem2014}
D.~Str{\"o}mbom, R.~Mann, A.~Wilson, S.~Hailes, A.~Morton, D.~Sumpter, and A.~King, ``Solving the shepherding problem: {{Heuristics}} for herding autonomous, interacting agents,'' \emph{Journal of The Royal Society Interface}, vol.~11, 2014.

\bibitem{haque2011biologically}
M.~A. Haque, A.~R. Rahmani, and M.~B. Egerstedt, ``Biologically inspired confinement of multi-robot systems,'' \emph{International Journal of Bio-Inspired Computation}, vol.~3, no.~4, pp. 213--224, 2011.

\bibitem{muro_wolf-pack_2011}
C.~Muro, R.~Escobedo, L.~Spector, and R.~Coppinger, ``\BIBforeignlanguage{en}{Wolf-pack ({Canis} lupus) hunting strategies emerge from simple rules in computational simulations},'' \emph{\BIBforeignlanguage{en}{Behavioural Processes}}, vol.~88, no.~3, pp. 192--197, 2011.

\bibitem{licitraSingleAgentIndirectherding2018}
R.~A. Licitra, Z.~I. Bell, E.~A. Doucette, and W.~E. Dixon, ``Single agent indirect herding of multiple targets: A switched adaptive control approach,'' \emph{IEEE Control Systems Letters}, vol.~2, no.~1, pp. 127--132, 2018.

\bibitem{li2023communication}
A.~Li, M.~Ogura, and N.~Wakamiya, ``Communication-free shepherding navigation with multiple steering agents,'' \emph{Frontiers in Control Engineering}, vol.~4, p. 989232, 2023.

\bibitem{piersonControllingNoncooperativeHerds2018}
A.~Pierson and M.~Schwager, ``Controlling {{Noncooperative Herds}} with {{Robotic Herders}},'' \emph{IEEE Transactions on Robotics}, vol.~34, no.~2, pp. 517--525, 2018.

\bibitem{zhang2024distributed}
S.~Zhang, X.~Lei, M.~Duan, X.~Peng, and J.~Pan, ``A distributed outmost push approach for multirobot herding,'' \emph{IEEE Transactions on Robotics}, vol.~40, pp. 1706--1723, 2024.

\bibitem{lien2004shepherding}
J.-M. Lien, O.~B. Bayazit, R.~T. Sowell, S.~Rodriguez, and N.~M. Amato, ``Shepherding behaviors,'' in \emph{IEEE International Conference on Robotics and Automation}, vol.~4, 2004, pp. 4159--4164.

\bibitem{aulettaHerdingStochasticAutonomous2022}
F.~Auletta, D.~Fiore, M.~J. Richardson, and M.~{di~Bernardo}, ``Herding stochastic autonomous agents via local control rules and online target selection strategies,'' \emph{Autonomous Robots}, vol.~46, no.~3, pp. 469--481, 2022.

\bibitem{lamaShepherdingControlHerdability2024}
A.~Lama and M.~di~Bernardo, ``Shepherding and herdability in complex multiagent systems,'' \emph{Physical Review Research}, vol.~6, p. L032012, 2024.

\bibitem{koAsymptoticBehaviorControl2020}
D.~Ko and E.~Zuazua, ``Asymptotic behavior and control of a ``guidance by repulsion'' model,'' \emph{Mathematical Models and Methods in Applied Sciences}, vol.~30, no.~04, pp. 765--804, 2020.

\bibitem{deMedioRobot1991}
C.~De~Medio and G.~Oriolo, ``Robot obstacle avoidance using vortex fields,'' in \emph{Advances in Robot Kinematics}.\hskip 1em plus 0.5em minus 0.4em\relax Springer, 1991, pp. 227--235.

\bibitem{geopf2024}
Y.~Gong, R.~Laha, and L.~Figueredo, ``Geopf: Infusing geometry into potential fields for reactive planning in non-trivial environments,'' in \emph{arXiv:2505.19688}, 2025.

\bibitem{vanhavermaet2024reactive}
S.~Van~Havermaet, Y.~Khaluf, and P.~Simoens, ``Reactive shepherding along a dynamic path,'' \emph{Scientific Reports}, vol.~14, p. 14915, 2024.

\bibitem{hamandi2024robotic}
M.~Hamandi, F.~Khorrami, and A.~Tzes, ``Robotic shepherding in cluttered and unknown environments using control barrier functions,'' in \emph{arXiv:2407.15701}, 2024.

\bibitem{chipade2021multiagent}
V.~S. Chipade and D.~Panagou, ``Multiagent planning and control for swarm herding in 2-d obstacle environments under bounded inputs,'' \emph{IEEE Transactions on Robotics}, vol.~37, no.~6, pp. 1956--1972, 2021.

\bibitem{zhi2021learning}
J.~Zhi and J.-M. Lien, ``Learning to herd agents amongst obstacles: Training robust shepherding behaviors using deep reinforcement learning,'' \emph{IEEE Robotics and Automation Letters}, vol.~6, no.~2, pp. 4163--4168, 2021.

\bibitem{khatib1986real}
O.~Khatib, ``Real-time obstacle avoidance for manipulators and mobile robots,'' \emph{The International Journal of Robotics Research}, vol.~5, no.~1, pp. 90--98, 1986.

\bibitem{sebastian_adaptive_2022}
E.~Sebastián, E.~Montijano, and C.~Sagüés, ``Adaptive {Multirobot} {Implicit} {Control} of {Heterogeneous} {Herds},'' \emph{IEEE Transactions on Robotics}, vol.~38, no.~6, pp. 3622--3635, 2022.

\bibitem{albi2016}
G.~Albi, M.~Bongini, E.~Cristiani, and D.~Kalise, ``Invisible control of self-organizing agents leaving unknown environments,'' \emph{SIAM Journal on Applied Mathematics}, vol.~76, no.~4, pp. 1683--1710, 2016.

\bibitem{olfatiFlocking2006}
R.~Olfati-Saber, ``Flocking for multi-agent dynamic systems: algorithms and theory,'' \emph{IEEE Transactions on Automatic Control}, vol.~51, no.~3, pp. 401--420, 2006.

\bibitem{Siciliano2009}
B.~Siciliano, L.~Sciavicco, L.~Villani, and G.~Oriolo, \emph{Robotics: Modelling, Planning and Control}.\hskip 1em plus 0.5em minus 0.4em\relax Springer, 2010.

\end{thebibliography}

\appendix
\renewcommand{\thetable}{A\arabic{table}}
\setcounter{table}{0} 

\label{appendix:numerical}
This appendix summarizes the key numerical settings used in the study.

\paragraph{Model parameters}
Table \ref{tab: parameters} reports the physical constants defining herder and target dynamics (Section \ref{sec: mathematization}). 

\begin{table}[H]
\centering
\vspace{4pt}
\caption{Model parameters for herders, targets, and environment used in the simulation study}
\label{tab: parameters}
\resizebox{\columnwidth}{!}{%
\begin{tabular}{ccc}
\toprule
Parameter & Notation & Value \\
\midrule
Herder's maximum velocity & $v_h$ & 7.5 m/s\\
Attraction strength to a selected target & $\alpha$ & 5\\
Offset distance from selected target & $\delta$ & 1.25~m\\
Balancing factor of obstacle avoidance for herders & $\gamma$ & 0.3\\
Influence radius of targets & $\lambda$ & 2.5~m\\
Strength of the repulsive drift from sensed herders & $\beta$ & 3 \\
Obstacle safety margin & $\epsilon_o$ & 1~m \\
Noise diffusion & $D$ & 0.5 \\
Domain radius & $\rho_0$ & 50~m \\
Goal region radius & $\goalradius$ & 10~m \\
Obstacle influence radius & $\lambda_o$ & 2.5~m\\
Repulsion strength from obstacles & $k_o$ & 10 \\
Sampling time & $\samplingtime$ & 0.01~s\\   
\bottomrule
\end{tabular}%
}
\end{table}

\paragraph{ROS~2 implementation details}

To handle Vicon tracking data and control robot velocities, we employ the ROS~2 environment.  
A \texttt{ros\_vicon\_bridge} node on the central computer publishes Vicon measurements in real time to a dedicated topic for each subject, denoted as  
\texttt{\textbackslash vicon\textbackslash subject},  
where \texttt{subject} corresponds to the namespace of each robot.  

A \texttt{controller} node subscribes to these topics, extracts the poses of all entities, and uses them as inputs to the control law. Based on the estimated positions of robots and obstacles, this node applies the dynamics in Eq.~\eqref{eq:herders_dynamics} for \textit{TurtleBot4} units and Eq.~\eqref{eq: target_dynamics} for \textit{Osoyoo} robots, accounting for their physical bodies and dynamics. The resulting linear and angular velocity commands are then published to the corresponding ROS~2 topics for execution.  
Table \ref{tab: parameters_exp} reports the physical constants defining herder and target dynamics during experiments (Section \ref{sec: exp}).

\begin{table}[H]
\centering
\caption{Model parameters for herders, targets, and environment used in the experiments}
\label{tab: parameters_exp}
\resizebox{\columnwidth}{!}{%
\begin{tabular}{ccc}
\toprule
Parameter & Notation & Value \\
\midrule
Attraction strength to a selected target & $\alpha$ & 3\\
Offset distance from selected target & $\delta$ & 0.375~m \\
Balancing factor of obstacle avoidance for herders & $\gamma$ & 0.3\\
Orbiting tangential gain & $\alpha_o$ & 4.5 \\
Orbiting radial gain & $\alpha_r$ & 3 \\
Threshold distance for orbiting & $r_{th}$ &  0.375~m\\
Offset distance for orbiting blend & $\epsilon_h$ & 0.1~m \\
Orbiting activation angle & $\beta_{th}$ & $\frac{pi}{4}$ rad \\
Orbiting deadband angle & $\beta_{orb}$ & $\frac{pi}{18}$ rad\\
Repulsion strength between robots of the same type & $k_{d}$ & 1 \\ 
Distance threshold between robots of the same type & $d_{ij}$ & 0.45~m \\ 
Influence radius of targets & $\lambda$ & 0.5~m \\
Strength of the repulsive drift from sensed herders & $\beta$ & 3 \\
Obstacle safety margin & $\epsilon_o$ & 0.1~m \\
Noise diffusion & $D$ & 0 \\
Obstacle influence radius & $\lambda_o$ & 0.2~m\\
Goal region radius & $\goalradius$ & 0.35~m \\
Repulsion strength from obstacles & $k_o$ & 10 \\
Sampling time & $\samplingtime$ & 0.002~s\\   

\bottomrule
\end{tabular}}%
\end{table}

\end{document}